\documentclass[aps,showpacs,preprintnumbers,amsmath,amssymb]{revtex4}
\usepackage{graphicx,amssymb,amsmath,subfigure}
\begin{document}
\preprint{}
\title{Electronic properties of closed cage nanometer-size spherical
graphitic particles}
\author{Godfrey Gumbs$^1$, Antonios  Balassis$^2$, Andrii Iurov$^1$,
 and Paula  Fekete$^3$}
\affiliation{$^1$Department of Physics and Astronomy,
Hunter College at the City University of New York, \\
695 Park Avenue New York, NY 10065, USA }

\affiliation{$^2$Physics Department, Fordham University,     441 East Fordham Road,
       Bronx, NY 10458-5198, USA }

\affiliation{$^3$West Point Military Academy, West Point, NY}
\date{\today}

\begin{abstract}
We  investigate the  localization
of charged particles by the image potential of
spherical shells, such as   fullerene
buckyballs.  These ''spherical image
states'' exist within   surface potentials formed
 by the competition between the attractive  image potential
 and the repulsive centripetal force arising
from the  angular motion.  The image potential has a power
law  rather than a logarithmic behavior for a nanotube,
leading to fundamental differences in the forms for the effective
potential for the two geometries. The sphere has localized stable
states close to its surface. At low temperatures, this results
in long lifetimes  for the image states. We predict the
possibility of creating image states  with binding energies of a few meV
 around metallic/non-metallic spherical shells
by photoionization. Applications and related phenomena are discussed.

\end{abstract}

\maketitle

   The experimental and theoretical study of carbon is currently
   one of the most prevailing research areas in condensed
   matter physics. Forms of carbon include several allotropes such as
   graphene,   graphite as well as the fullerenes, which cover
	any molecule composed entirely of carbon, in the form of
a hollow sphere, ellipsoid or tube. Like graphite, fullerenes
are  composed of stacked graphene sheets of linked hexagonal rings.
For these, the carbon atoms
form strong covalent bonds through hybridized   sp$^2$ atomic
orbitals between three nearest neighbors in a planar or nearly planar
configuration.

Mass spectrometry experiments showed strong peaks corresponding
to molecules with the exact mass of sixty carbon atoms and
other carbon clusters such as C$_{70}$, C$_{76}$, and up to
C$_{94}$ \cite{1,1b}.
Spherical fullerenes,  well known as ``buckyballs" (C$_{60}$),
were   prepared in 1985 by  Kroto,  et al. \cite{2}
 The structure was also identified
 about five years earlier by  Iijima, \cite{3}  from an electron microscope
image, where it formed the core of a multi-shell fullerene or
``bucky onion."  Since then, fullerenes
have  been found to exist  naturally \cite{4}. More recently,
fullerenes have been detected in outer space \cite{5}.
As a matter of fact, the discovery of fullerenes greatly
expanded the number of known
carbon allotropes, which until recently were limited to
graphite, diamond, and amorphous carbon such as soot and charcoal.
Both buckyballs and carbon nanotubes, also referred to as buckytubes
have been the focus of intense
investigation,  for their unique chemistry as well as   their technological
applications in materials science, electronics, and
nanotechnology \cite{6}.

Recently, the image states of metallic carbon nanotubes \cite{7}
and double-wall non-metallic nanotubes  \cite{8,8b} were investigated
Experimental work \cite{8c} includes photoionization \cite{8c+}
and time-resolved photoimaging of image-potential states in
carbon nanotubes \cite{8cc+}.
There  has been general interest \cite{8c++} in these
structures because  of electronic control on the nanoscale using
image states. This has  led to wide-ranging potential applications
including field ionization of cold atoms near carbon nanotubes,
\cite{8d}   and chemisorption of  fluorine atoms on the surface of
carbon nanotubes \cite{8e}.
Here,  we calculate the nature of the image-potential states in
a spherical   electron gas (SEG) confined to the  surface
of a buckyball in a similar fashion as in the case of a nanotube.
For the semi-infinite metal/vacuum
interface, \cite{9}  related image-potential
states  have been given a considerable amount of theoretical
 attention over the years.  Additionally, these
 states have been observed for pyrolytic graphite \cite{10}
and metal supported graphene \cite{11}. Silkin,   et al.  \cite{12}
further highlighted the importance of image  states in
graphene \cite{13} by concluding that
the inter-layer state in graphite is formed by the
hybridization of the lowest image-potential state in graphene
in a similar way as  occurs in  bilayer graphene \cite{14,14a}.  The
significance of the role played by  the
image potential has led to  the observation  that
for planar layered materials,
strongly dispersive inter-layer states are in common. However,
the eigenstates for a spherical shell are non dispersive and so
too are the collective plasma modes, \cite{15,15a,16,16a,16b}
leading to interesting properties for the image potential.
Fullerene is an unusual reactant in many organic reactions such
as the Bingel reaction that allows the attachment of extensions
to the fullerene \cite{17}.  For this reaction,
the shorter bonds located at the intersection
 of two hexagons (6-6 bonds) are the
preferred double bonds  on the
fullerene surface.  The driving force is
 the release of steric strain. The surface-related properties
of these structures are of course influenced by its image potential,
making it one reason why we are interested in this potential.
\begin{figure}[ht!]
\centering
\includegraphics[width=0.4\textwidth]{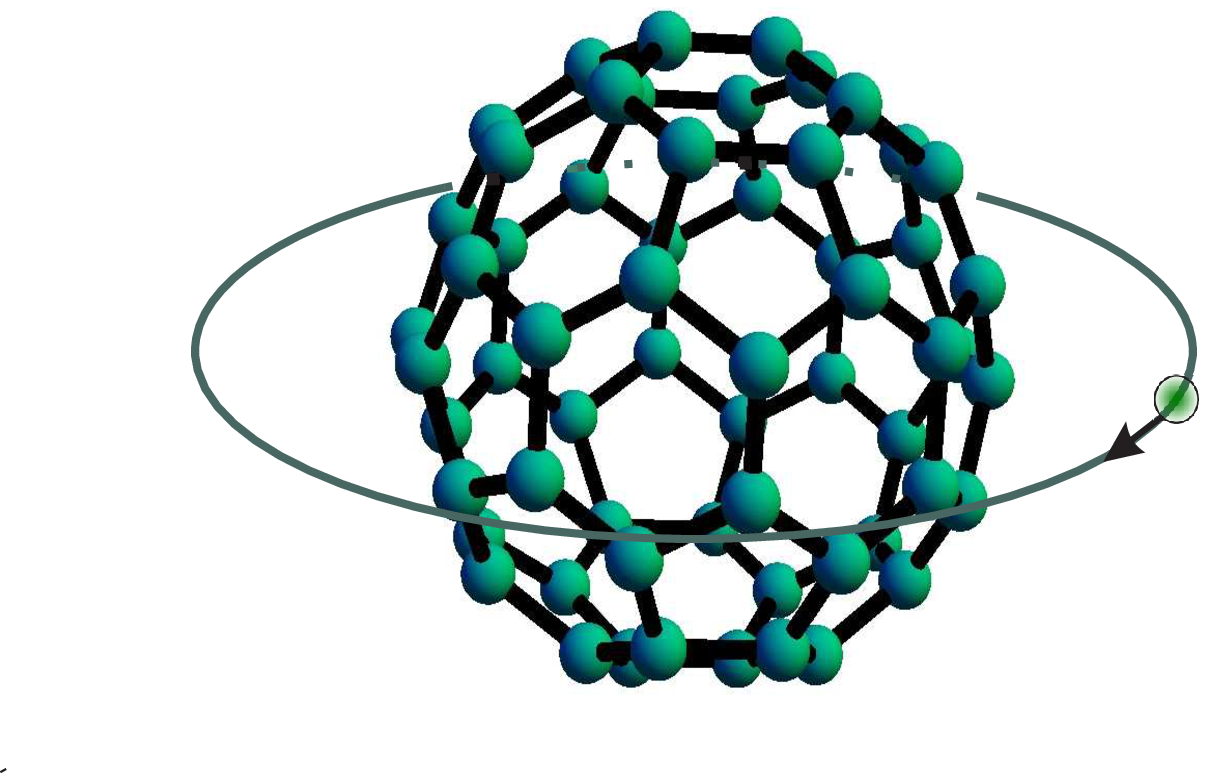}
\caption{(Color online) Schematic illustration of a charged particle
captured by the image potential and orbiting  around a   buckyball. The radius of the
orbit is determined by the dielectric constant within and surrounding
the shell as well as the angular momentum quantum number of the
captured particle.  For semiconducting shells, the localization
is strong and the radius of the stable orbit can be a few nanometers.
The localization is weak for metallic shells.}
\label{FIG:1}
\end{figure}

The fullerenes are modeled in analogy with the collective
excitations in a planar semiconductor  two-dimensional-electron-gas.
We now consider a spherical shell of radii $R$    whose center is at
the origin. The background dielectric constant is $\epsilon_1$
for $0<r<R$,  and $\epsilon_2$ for $r>R$. An electron gas is
confined to the surface of   the sphere. If a charge $Q$
is located at $(r_0,\theta_0,\phi_0)$ in spherical coordinates,
then for $r_0>R$, the total electrostatic potential is given by $\Phi_{\rm{tot}}=\Phi_{\rm{ext}}+\Phi_{\rm{ind}}$, where
$\Phi_{\rm{ext}}$ is the external potential due to the point particle and  $\Phi_{\rm{ind}}$ is the induced potential. The external potential
 may be expanded in the form

\begin{equation}
\Phi_{\rm{ext}}\left(r,\theta,\phi\right)
=4\pi kQ\sum_{LM}\frac{1}{2L+1}
\frac{r_<^L}{r_>^{L+1}}
Y_{LM}^{\ast}\left(\Omega_0\right)Y_{LM}\left(\Omega\right)\ ,
\end{equation}
where $k=(4\pi\epsilon_0)^{-1}$ with $\epsilon_0$ the permittivity
of free space. Also,  $Y_{LM}\left(\Omega\right)$ is a spherical harmonic and
$\Omega$ is a solid angle. For $r<R$ we express the total potential in
the form

\begin{equation}
\Phi_{\rm{tot}}^{\left(1\right)}\left(r,\theta,\phi\right)
=4\pi kQ\sum_{LM}\frac{1}{2L+1}A_{L}r^{L}
Y_{LM}\left(\Omega\right),
\label{potential1}
\end{equation}
whereas for $r>R$ we express the induced potential as

\begin{equation}
\Phi_{\rm{ind}}^{\left(2\right)}\left(r,\theta,\phi\right)
=4\pi kQ\sum_{LM}\frac{1}{2L+1}
B_{L}r^{-\left(L+1\right)}Y_{LM}\left(\Omega\right)\ .
\label{inducedp}
\end{equation}
The total potential $\Phi_{\rm tot}^{(2)}(r,\theta,\phi)
=\Phi_{\rm ext}(r,\theta,\phi)+\Phi_{\rm ind}^{(1)}(r,\theta,\phi)$
for $r>R$ then becomes

\begin{eqnarray}
&&\Phi_{\rm{tot}}^{\left(2\right)}\left(r,\theta,\phi
\right)=4\pi kQ\sum_{LM} \frac{1}{2L+1}
\nonumber\\
&\times&
\left[\frac{r_<^L}{r_>^{L+1}}
Y_{LM}^{\ast}\left(\Omega_0\right)+B_L r^{-\left(L+1\right)}\right]Y_{LM}\left(\Omega\right)\ .
\label{potential2}
\end{eqnarray}

On the surface of the sphere, the boundary conditions are
$\Phi_{\rm{tot}}^{\left(1\right)}\left(R,\theta,\phi\right)
=\Phi_{\rm{tot}}^{\left(2\right)}\left(R,\theta,\phi\right)
$ and
$
\left.
\left[\epsilon_1\Phi_{\rm{tot}}^{\left(1\right)}
\left(r,\Omega;\omega\right)-\epsilon_2 \Phi_{\rm{tot}}^{\left(2\right)}\left(r,\Omega;\omega
\right)\right]\right|_{r=R}^\prime =4\pi k\sigma\left(R,\theta,
\phi;\omega\right)\ ,
$
where $\sigma\left(R,\theta,\phi;\omega\right)$ is the induced
surface charge density on the spherical shell. Expanding
$\sigma$ in terms of spherical harmonics and using linear
response theory we find

\begin{equation}
\sigma\left(R,\theta,\phi;\omega\right)=-\frac{2kQe^2}{R^2}
\sum_{L M}\frac{1}{2L+1}A_L R^L
\Pi_L(\omega)Y_{LM}\left(\Omega\right)\ .
\label{density}
\end{equation}
In this equation, $\Pi_L(\omega)$ is the SEG polarization
function  for $L$ an integer  and given in terms of the
Wigner $3$-$j$ symbol \cite{15,16}

\begin{equation}
\Pi_L(\omega) = \sum_{\ell \ell^{\prime}}
\frac{f_0(E_\ell)-f_0(E_{\ell^\prime})}
{\hbar\omega+E_{\ell^\prime}-E_\ell}\
(2\ell+1)(2\ell^\prime+1)
\left( \begin{matrix}
l&l^\prime& L\cr
 0 & 0 & 0\cr
\end{matrix}\right)^2 \ ,
\label{polarization}
\end{equation}
where $f_0(E)$ is the Fermi-Dirac distribution function
and $E_\ell=\hbar^2\ell(\ell+1)/(2m^\ast R^2)$
with $\ell=0,1,2,\cdots$ and $m^\ast$ is the electron effective
mass.   The induced potential
$\Phi_{\rm{ind}}^{\left(2\right)}\left(\bf r ;\omega\right)$
outside the spherical shell may be calculated to be

\begin{eqnarray}
\Phi_{\rm{ind}}^{\left(2\right)}\left(\bf r ;\omega\right)
&=&4\pi kQ\sum_{L}\left[\frac{\epsilon_2}{\varepsilon_L(\omega=0)}
-\frac{1}{2L+1}\right]\frac{R^{2L+1}}{r_0^{L+1}}
\nonumber\\
&\times& \frac{1}{r^{L+1}} \sum_{M}Y_{LM}^{\ast}\left(\Omega_0\right)Y_{LM}\left(\Omega\right).
\end{eqnarray}
where
$
\varepsilon_L(\omega)=L \left(\epsilon_1+\epsilon_2\right)
+\epsilon_2+ (2e^2/R)\Pi_{L}(\omega)$
is the dielectric function of the SEG.
The force on a  charge $Q$ at ${\bf r}_0=(r_0,\phi_0,\theta_0)$
is along the radial direction and can be found using
$F(r_0)=-Q\left.
\displaystyle \partial \Phi_{\rm{ind}}^{\left(2\right)}({\bf{r}})/\partial r
\right|_{{\bf{r}}_0}$ yielding

\begin{equation}
F(r_0)=kQ^2 \sum_{L}(L+1)\left[(2L+1)
\frac{\epsilon_2}{\varepsilon_L(\omega=0)}-1\right]
\frac{R^{2L+1}}{r_0^{2L+3}} \ .
\label{Fr0}
\end{equation}
The interaction potential energy ${\cal U}_{\rm{im}}(r_0)$
may now be calculated from

\begin{eqnarray}
&&{\cal U}_{\rm{im}}(r_0)
\equiv  \sum_L {\cal U}_{\rm{im}}^{(L)}(r_0)=\frac{1}{2}Q\Phi_{\rm ind}(r_0,\theta_0,\phi_0)
\nonumber\\
&=& \frac{kQ^2}{2r_0}
\sum_{L}\left[(2L+1)\frac{\epsilon_2}
{\varepsilon_L(\omega=0)}-1\right]\left(\frac{R}{r_0}\right)^{2L+1}
\end{eqnarray}

\begin{figure}[ht!]
\centering
\includegraphics[width=0.36\textwidth]{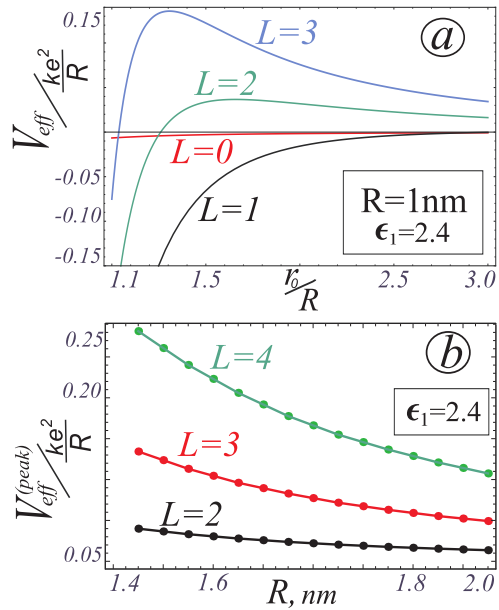}
\caption{\label{FIG:2} (Color online) (a) The effective
potential    $V_{\rm eff}$  between
a charged particle and a spherical  shell
is shown for a number of
angular momenta $L$.   The radius of the sphere is $R=1$ nm
 and we chose $\epsilon_1 =2.4,\ \epsilon_2 =1$.  In (b),
the height of the peak in the effective potential appearing in
(a) is plotted as a function of the radius.}
\end{figure}

\begin{figure}[ht!]
\centering
\includegraphics[width=0.37\textwidth]{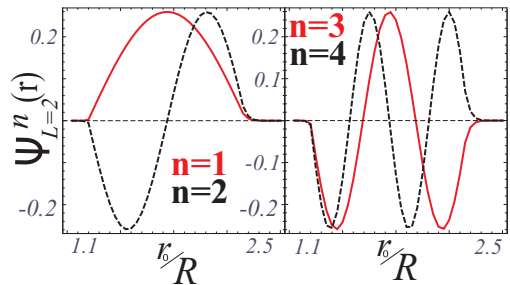}
\caption{(Color online) The  wave functions for the
ground state ($n=1$) and first three excited states ($n=2,3,4$)
are plotted for the effective potential $V_{\rm eff}$ between
a charged particle and a spherical  shell  when  $L=2$.
We chose $\epsilon_1 =2.4,\ \epsilon_1 =1$ and
$R=1\; nm$}
\label{FIG:3}
\end{figure}

 The effective potential is the sum of the image potential
 and the centrifugal term
and is given by \cite{7,8}

\begin{equation}
V\ _{\mbox{eff}}^{(L)}(r_0,\theta)=  {\cal U}_{\rm{im}}^{(L)}(r_0)
+\frac{\hbar^2\left(L^2+L-\frac{1}{4}\right)}{2M^\ast (r_0\sin\theta_0)^2}
 \ ,
\label{v-eff}
\end{equation}
showing that $V_{\mbox{eff}}^{(L)}$ is not spherically symmetric.
In this notation,   $M^\ast$ is the mass of the
captured charged particle in
an orbital state with angular momentum quantum number $L$.
In Fig. \ref{FIG:2}(a),
we calculated $V_{\mbox{eff}}^{(L)}(r_0,\theta)$ as a function of $r_0$,
for chosen $L,\ R$. Also,  for the background dielectric constant, we
chose $\epsilon_1=2.4$, corresponding to graphite,  and
$\epsilon_2=1$. The electron effective mass used in calculating
the polarization function $\Pi_L$  in Eq.\ (\ref{polarization})
was $m=0.25 m_e$ where $m_e$ is the bare electron mass,
and the Fermi energy $E_F=0.6\ eV$.
The orbiting particle effective mass $M^\ast=m_e$.
Figure \ref{FIG:2}(b)  shows how the peak values of the
effective potential depend on  radius. Of course, the
height of the peak is linked to the localization of
the particle in orbit. In Fig. \ref{FIG:3},
the ground and three lowest excited state wave functions
are plotted for the effective
potential $V_{\rm eff}^{(L)}$ when $L=2$ in Fig. \ref{FIG:2}(a).

The value of the angular momentum quantum number $L$ as well as
the curvature of the surface of these complex carbon structures
clearly plays a crucial role in shaping the effective potential.
Generally, the form for the $L$-th term may be expressed as
$V_{\rm eff}^{(L)}(r_0)=-\alpha_L  r_0^{-2(L+1)}+\beta_L r_0^{-2}$,
where $\alpha_L$ and $\beta_L$ are due to the image potential
and centrifugal force, respectively. The coefficient $\alpha_L$
is always positive whereas $\beta_L$ is only negative for $L=0$.
This power-law behavior ensures that no matter what values
the two coefficients may have the image term dominates  the
centrifugal term, leading to a local maximum in the effective
potential. This is unlike the behavior for a cylindrical
nanotube where the image term is logaritmic,
 due to the {\em linear} charge distribution, and may be
dominated by the $r_0^{-2}$ centrifugal term, leading
to a local minimum instead. Consequently, capturing and
localizing a charged particle  by the image potential
of spherical conductors and dielectrics is fundamentally different
from that for a cylindrical nanotube. For the sphere, as shown in Fig.\
\ref{FIG:3}, the wave function is more localized around the
spherical  shell within its effective potential, i.e., the wave
function is not as extended . Additionally, the confinement of the
charged particle is close to the spherical surface.

\begin{figure}
 \centering
\subfigure[]{
\includegraphics[width=0.49\textwidth]{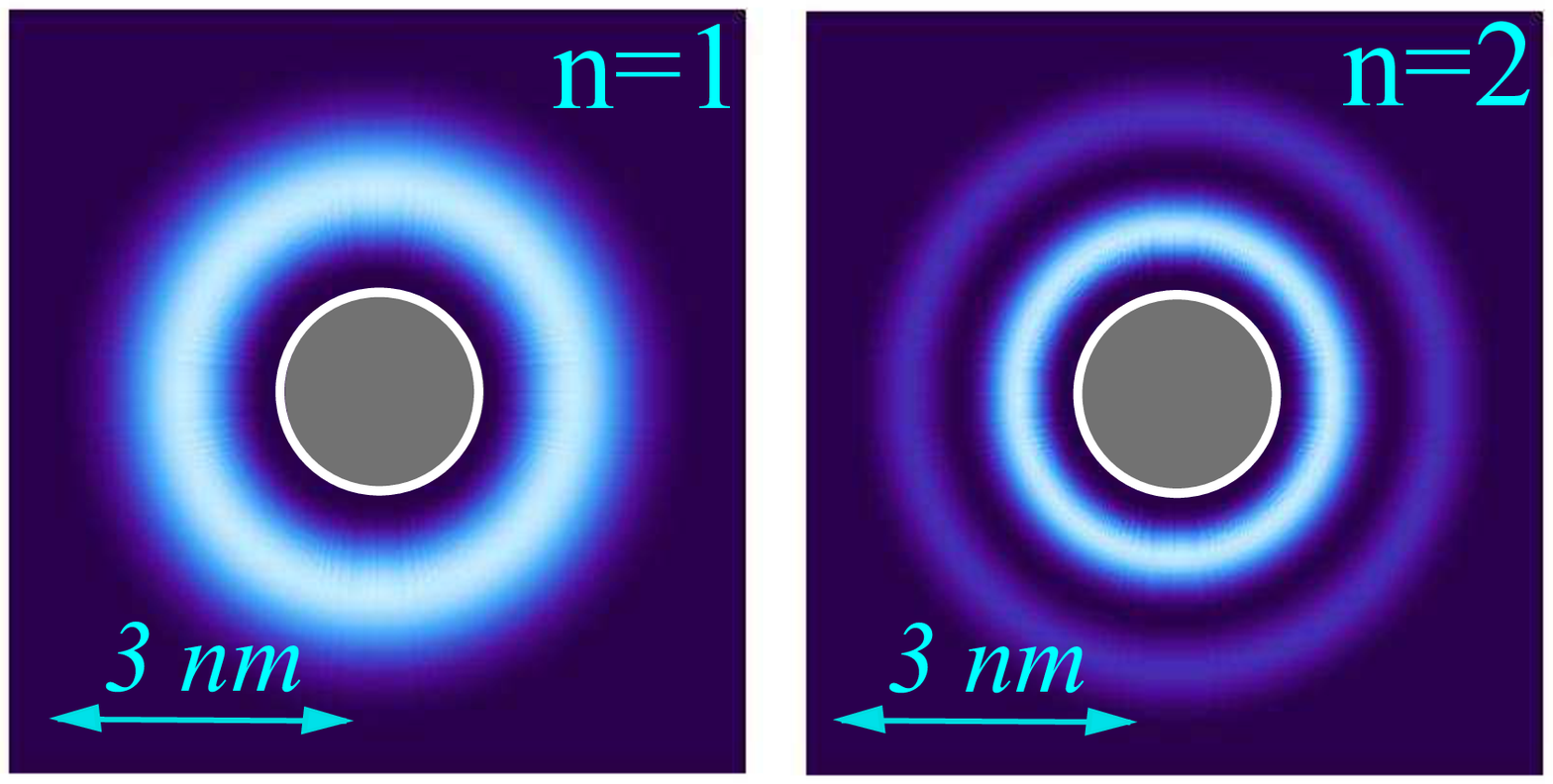}
}
\subfigure[]{
\includegraphics[width=0.49\textwidth]{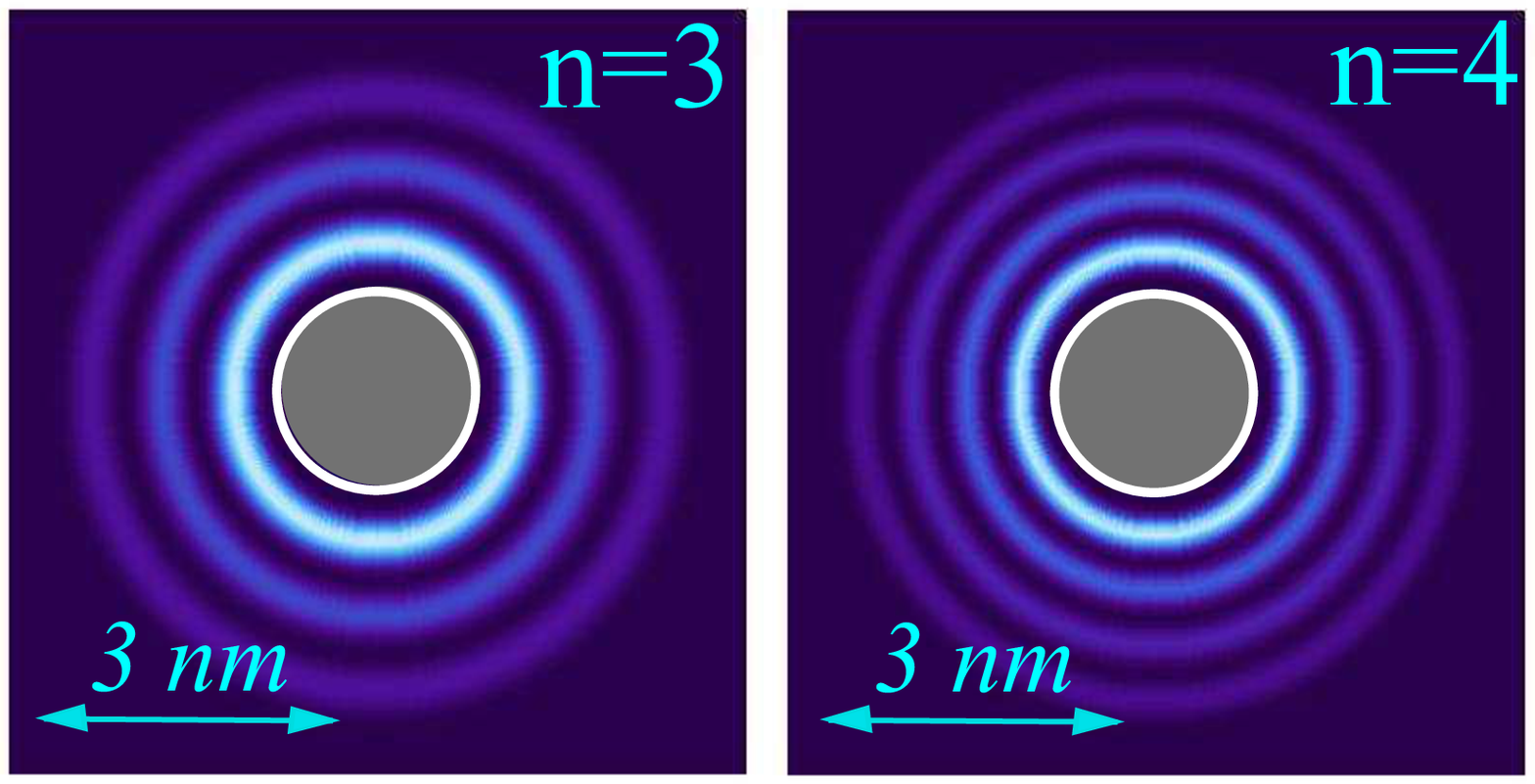}
}
\caption{(Color online) Probability density plots for
$|\Psi_{L,n}(r_0)|^2/r_0^2$ when $L=2$ and $n=1,2$ (upper panel)
as well as $n=3,4$ (lower panel), where $n$ labels the eigenstates,
  for a spherical shell of radius $R=10$\ \AA. The wave function $\Psi_{L,n}(r_0)$
is a solution of the one-dimensional Schr\"odinger equation with effective
potential $V_{\mbox{eff}}(r_0,\theta=\pi/2)$ shown in Fig.\
\ref{FIG:2}. We chose $\epsilon_1=2.4$
inside the ball, whose outline is shown as a thin circle,
 and $\epsilon_2=1$ in the surrounding medium.}
\label{FIG:4}
\end{figure}

The choice for the radius  does  render   some crucial
changes in  $V_{\rm eff}^{(L)}$ to make a  difference
in the location and height of the  peak. However,
only the higher-lying localized states are affected.
In Fig. \ \ref{FIG:2}, we show how the peak height changes as $L$
 is varied. Additional   numerical results
corresponding to $\epsilon_1\gg \epsilon_2$ have shown  that a
spherical metallic shell has a reduced potential peak for confining
the captured charge.   Thus, the spherical metallic
 shell is not as susceptible   for particle confinement
in highly excited states as the
metallic nanotube. \cite{7,8,8b}  This
indicates that the  dimensionality plays a non-trivial role in
 formation  of image states and their spatial extension near the
surface of the nanometer-size graphitic structure.    This
direct crossover from a one-dimensional to a three-dimensional
regime is not determined by  polarization effects for the structure
in the metallic limit since in the limit $\epsilon_1\to\infty$
 in Eq.\ (\ref{Fr0}), the
$\Pi_L(\omega)$  term makes no contribution. The difference is
due entirely  to the geometrical shape in the metallic regime
where  graphitic plasmons fail to develop.  For finite values
of $\epsilon_1$,    we encounter the  regime where
excited particles contribute through the polarization function
$\Pi_L(\omega)$ defined in Eq.\ (\ref{polarization}).
 The behavior of plasmon excitation as a function of angular
 momentum quantum number $L $ for fullerenes
resembles in all respects the long wavelength ($q\to 0$) limit
 of carbon nanotubes \cite{18,19,20}.      Furthermore, in
the case of the low-frequency $\pi$- plasmons in carbon nanotubes,
a  surface mode may develop for large $q$, due to the
difference in the values for the dielectric constants within
the graphitic structure and the surrounding medium \cite{21}.

Since the polarization function $\Pi_L(\omega)$ vanishes identically
for $L=0$,  the attractive part of the effective potential is only
significantly modified by screening for a fast-rotating
electron. This behavior at zero angular momentum differs from
tubular-shaped image states for  single-walled carbon
nanotubes which are formed in a potential isolated from
the tube \cite{7}.  The large   angular momentum   image states
for spheres may be  probed by  femtosecond time-resolved
photoemission \cite{8cc+}. Our formalism shows that
considering photoionization from various levels of C$_{60}$,
the Coulomb interaction between an external charge and its
image is screened by the statically stretched SEG
through the  dielectric function $\epsilon_L(\omega=0)$.
The polarization of the medium $\Pi_L$ which is driven
by the electrostatic interaction  is generated by particle-hole
transitions across the Fermi surface.  The polarization
also determines the Ruderman-Kittel-Kasuya-Yosida (RKKY)
interaction energy between two magnetic impurities as well
as the induced  spin density due to a magnetic impurity.

Increasing  radius, the position of the
 peak  moves closer to the  sphere  as $R^{-1/(2L)}$.
 In the absolute units, the position of the peak depends as
$r_0^{(peak)} \backsim R^{1-1/(2L)}$.
The typical distances from the surface are between $1.3\ R$
and $1.5\ R$.  Figure\ \ref{FIG:2} (b) demonstrates how the potential peaks
 (corresponding to the local maximum for the $V_{\rm eff}$)
depend on the radius of the buckyball for various
angular momentum quantum number  $L$.  Clearly,
we see that the potential  peak decreases with increased
 radius leading us to  conclude that confinement is
strongest for smaller buckyballs and particles with
large angular momentum. For the nanotube, increasing
$L$ leads to a reduced local minimum in the effective potential
and the ability to localize the charge \cite{7}.
Approximately, the curves may be fitted
analytically to $ \backsim 1/R$ for all considered values of
 $L$. However, we found that a better  fit for
$L>5$  would be of the form  $c_1/R+c_2/R^2$ where
$c_1  , c_2$ are constants.

\par

For increased $\epsilon_1$, i.e., the metallic limit with
$\epsilon_1 \gg \epsilon_2$, we have
$\epsilon_L \backsimeq  \epsilon_1 L$,
 so that the coefficient
$\left( (2L+1)[\epsilon_2/\epsilon_L]  -1 \right)  \to -1$.
In the case of  dielectric constant $\epsilon_1\sim 2.4$ for
the buckyball, the above mentioned coefficients lie  within
 the range from $-0.4$ to $-0.9$ and decreases  with increasing $L$.
Consequently, for the transition to the metallic limit, these
coefficients  are more affected for states  with large $L$. So,
we may conclude that $\epsilon_1$ has little effect on the
position and  height of the peak in the effective potential.
In fact, for the metallic case, the peak is observed to be slightly
further away  from the center (very little difference $\sim 1.37\ R$
compared to $\sim 1.31\ R$ for $R=1\; nm$). The height  of the
peak is only slightly decreased in the metallic case ($0.24$ compared
to $0.27$ in the case of fullerenes). These numbers are provided
for fixed $L$ and the unit of energy is the same as that in Fig.\ \ref{FIG:2}.

\par
Regarding the wave functions and density plots,
Figs.\ \ref{FIG:3}, and \ref{FIG:4}  demonstrate
the wave function of a bounded electron trapped between the infinite
hard wall of the sphere and the potential peak. First, we note
that we obtained qualitatively similar behavior for different values
of $L$, so the electron states  corresponding to  the potentials
with different angular momenta are almost the same. We clearly see
that the electron wave functions are not exactly localized in the
"potential well" due to the asymmetry of the boundary conditions, i.e.,
infinitely high wall on the left and the effective potential
profile on the right-hand side.  The wave functions corresponding
to $L=0$ are extremely delocalized  due to the relatively shallow
 potential.  The fact that the effective potential is not spherically
symmetric means that for arbitrary angle $\theta$, we must solve
a three-dimensional Schr\"odinger equation.  However, for 
trajectories parallel to the $x-y$  plane when the angle $\theta$ 
is a constant of motion,  the problem reduces to a quasi-one-dimensional
Schr\"odinger equation involving the radial coordinate. 
In our calculations, we set $\theta=\pi/2$ so  that the captured 
charge is moving in the equatorial plane.  For this, the centrifugal 
term is weakest compared to the image potential, but still affords 
us the opportunity to see its effect on localization.  
The density plots in Fig.\ \ref{FIG:4}
show that the innermost ring is substantially brighter than the
outer rings. This is a consequence of the presence of the $r_0^{-2}$
factor in the electron probability function. In contrast, the
corresponding plots for the nanotube \cite{8b} do not have the
innermost ring so much brighter than the outer rings
 because of the fact that the density function
in that case depends on inverse distance of the charge from the center
of the cylinder instead.  This is another unusual specific
feature of the considered geometry and indicates that the captured
charge is more strongly localized for the spherical shell
closest to the surface for the sphere
than the cylindrical nanotube. The  lowest bound states for Fig.\ \ref{FIG:3}
are   in the range from $-10$ to  about$-100$ meV, with first
few excited states lying very close to the ground state energy.
Of course, the bound state energies may be adjusted by varying the
radius $R$ since our calculations have shown that the peak
potential decreases as a power function with  increasing radius.
This property allows considerable manipulation of a captured
electron and its release to a source of holes for recombination
and release of a single photon whose frequency and polarization are linked to the
electron.    This single-photon source
may have variable frequency with a broad range of applications
in quantum information
where the message is encoded in the number of photons transmitted
from node to node in an all-optical  network.
 Gate operations are performed by the nodes based on
quantum interference effects between  photons which cannot
be identified as being  different.   The low frequency photons could be in the
infrared which is most useful range for telecommunications.
Another, more general, practical application and technological
use of such unique
quantum states  would be to quantum   optical metrology of high-accuracy and
absolute optical measurements.

\acknowledgments
This research was supported by  contract \# FA 9453-07-C-0207 of
AFRL.




\begin{thebibliography}{0}
\expandafter\ifx\csname natexlab\endcsname\relax\def\natexlab#1{#1}\fi
\expandafter\ifx\csname bibnamefont\endcsname\relax
  \def\bibnamefont#1{#1}\fi
\expandafter\ifx\csname bibfnamefont\endcsname\relax
  \def\bibfnamefont#1{#1}\fi
\expandafter\ifx\csname citenamefont\endcsname\relax
  \def\citenamefont#1{#1}\fi
\expandafter\ifx\csname url\endcsname\relax
  \def\url#1{\texttt{#1}}\fi
\expandafter\ifx\csname urlprefix\endcsname\relax\def\urlprefix{URL }\fi
\providecommand{\bibinfo}[2]{#2}
\providecommand{\eprint}[2][]{\url{#2}}

\end{thebibliography}


\begin{references}

\bibitem{1}  E.  Osawa, Kagaku \textbf{25}, 854  (1970).

\bibitem{1b} Joseph F. Anacleto, Michael A. Quilliam,
Anal. Chem.  \textbf{65},   2236  (1993).

\bibitem{2}  H.W. Kroto, J. R. Heath, S. C. O'Brien, R. F. Curl,
and R. E. Smalley,    Nature \textbf{318}, 162  (1985);


\bibitem{3}  S Iijima,  Journal of Crystal Growth
\textbf{50}, 675 (1980).


\bibitem{4}   P.R. Buseck,  S.J. Tsipursky, R.  Hettich,
Science \textbf{257}, 215 (1992).


\bibitem{5}    J.  Cami, J.  Bernard-Salas,  E. Peeters, S. E.Malek,
Science \textbf{329}, 180   (2010).


\bibitem{6}     C. A. Poland,  et al., Nature Nanotechnology \textbf{3},
423  (2008).

\bibitem{7} Brian E. Granger, Petr Kr\'al, H.R. Sadeghpour, and
Moshe Shapiro, Phys. Rev. Lett. {\bf 89}, 135506 (2002).

\bibitem{8}   Godfrey Gumbs, Antonios Balassis, and Paula Fekete,
Phys. Rev. B \textbf{73}, 075411 (2006).


\bibitem{8b}     S. Segui, C. Celedon López, G. A. Bocan, J. L. Gervasoni,
 and N. R. Arista, Phys. Rev. B \textbf{85}, 235441 (2012).


\bibitem{8c}   K. Schouteden, A. Volodin, D. A. Muzychenko, M. P. Chowdhury,
A. Fonseca, J. B. Nagy, and C. Van Haesendonck,
Nanotechnology \textbf{21}, 485401 (2010).


\bibitem{8c+}   Matthew A. McCune, Mohamed E. Madjet, and
Himadri S. Chakraborty,   Journal of Physics B: Atomic, Molecular
and Optical Physics \textbf{41}, 201003 (2008).

\bibitem{8cc+}
 M. Zamkov, N. Woody, S. Bing, H. S. Chakraborty, Z. Chang,
 U. Thumm, and P. Richard,
Phys. Rev. Lett. \textbf{93}, 156803 (2004).

\bibitem{8c++}   Silvina Segui, Gisela A. Bocan, Néstor R. Arista,
and Juana L. Gervasoni, Journal of Physics: Conference Series
\textbf{194}, 132013 (2009).



\bibitem{8d}        Anne Goodsell, Trygve Ristroph, J. A. Golovchenko,
and Lene Vestergaard Hau, Phys. Rev. Lett. \textbf{104}, 133002 (2010).


\bibitem{8e}          V. l. .A. Margulis and E. E. Muryumin,
Physica B: Condensed Matter \textbf{390}, 134 (2007).



 \bibitem{9}    P. M. Echenique and J. B. Pendry, J. Phys.:
 Condens. Matter \textbf{11}, 2065 (1978).


\bibitem{10}    J. Lehmann, M. Merschdorf, A. Thon, S. Voll, and
W. Pfeiffer, Phys. Rev. B \textbf{60}, 17037 (1999).


\bibitem{11}     19 I. Kinoshita, D. Ino, K. Nagata, K. Watanabe,
N. Takagi, and Y.   Matsumoto, Phys. Rev. B \textbf{65}, 241402(R) (2002).


\bibitem{12}  V. M. Silkin, J. Zhao, F. Guinea, E. V. Chulkov,
 P. M. Echenique, and H. Petek, \prb  \textbf{80}, 121408(R) (2009).

\bibitem{13}  Godfrey Gumbs, D. Huang, and P. M. Echenique,
Phys. Rev. B \textbf{79},035410 (2009).

\bibitem{14}  J. Zhao, M. Feng, J. Yang, and H. Petek, ACS Nano
\textbf{3}, 853  (2009).

\bibitem{14a}   Lu-Jing Hou and Z. L. Miskovic,
Phys. Rev. E \textbf{77}, 046401 (2008).


\bibitem{15}  Takeshi Inaoka, Surface Science \textbf{273}, 191  (1992) .


\bibitem{15a}    P.   Longe,
 Solid State Communications   \textbf{97},   857  (1996).


\bibitem{16}  J. Tempere,  I. F. Silvera,  and J. T. Devreese,
\prb \textbf{65}, 195418(2002).


\bibitem{16a} J. Tempere, I.F. Silvera, J.T. Devreese,
Surface Science Reports \textbf{62},  159  (2007).


\bibitem{16b}   Constantine Yannouleas, Eduard N. Bogachek,
and Uzi Landman, \prb  \textbf{53},  10 225  (1996).

\bibitem{17}   Carsten  Bingel,   Chemische Berichte \textbf{126},
 1957 (1993).

\bibitem{18}   Godfrey Gumbs and G. R. Aizin, Phys. Rev. B 65,
195407 (2002).


\bibitem{19} M.F. Lin and W.K. Kenneth Shung, Phys. Rev. B \textbf{47},
6617 (1993).


\bibitem{20} M.F. Lin and W.K. Kenneth Shung, Phys. Rev. B
\textbf{48}, 5567 (1993).


\bibitem{21}   A. Bahari and A. Mohamadi, Nuclear Instruments
and Methods in Physics Research Section B: Beam Interactions with
Materials and Atoms  \textbf{268}, 3331 (2010).






\end{references}
\end{document}